\documentclass[prl,twocolumn,showpacs]{revtex4}
\usepackage{graphicx}
\newcommand{\feyn}[1]{{#1}\!\!\!{\slash}}
\newcommand{\bfone}{1\hspace{-1.3mm}1}

\begin{document}
 
\title{The Color-Superconducting {'}t Hooft Interaction}
 
\author{Andrew W. Steiner}
\affiliation{Theoretical Division, Los Alamos National Laboratory,
Los Alamos, NM  87545}
\date{\today}

\begin{abstract}
We consider the effect of a six-fermion interaction of the {'}t Hooft
form in the quark-quark channel on the ground state of matter at
finite density. The coupling constant for this new term is varied
within the limits suggested by naturalness. The flavor-mixing effects
of the additional term destabilize the color-flavor-locked (CFL) and,
to a lesser extent, the two-flavor color superconducting (2SC) phases of
quark matter, especially for positive values of the coupling. For
some values of the coupling, the critical density for CFL phase is
nearly larger than the maximum density in the neutron stars. We
comment on the implications for neutron star evolution.
\end{abstract}

\pacs{12.38.-t, 12.39.Fe, 26.60.+c, 97.60.Jd}
\maketitle

\section{History and Motivation}

The study of dense (baryon chemical potential $\sim$ 1.5 GeV) matter
has been recently revolutionized by the observation that dense quark
matter exhibits color-superconductivity and that the gaps may be of
order 100 MeV~\cite{Bailin84, Alford98, Rapp98}. Gaps of this
magnitude are large enough to have significant implications for neutron
star structure~\cite{Reddy03}, proto-neutron star
evolution~\cite{Carter00}, and neutron star cooling~\cite{Page00}.

Since directly utilizing QCD at the relevant densities
($\mu_{\mathrm{baryon}} \sim 1$ GeV) is so far impossible, the use of
effective theories like the Nambu--Jona-Lasinio (NJL) model is common
in the study of dense quark matter. In the NJL model, the high-energy
degrees of freedom (the gluons) are integrated out and we restrict
ourselves to working at energy scales less than the momentum 
cutoff $\Lambda$~\cite{Nambu61,Hatsuda94}
\begin{equation}
{\cal L}_{\mathrm{eff}} = \sum_n 
\frac{c_n}{\Lambda^{\mathrm{dim}({\cal O}_n)-4}} {\cal O}_n
\end{equation}
where ${\cal O}_n$ are operators, $\mathrm{dim}({\cal O}_n)$ is
the dimension of the operator, and $c_n$ are dimensionless
coupling constants. Because it is impossible to create a model
with the same symmetries as QCD with four-fermion operators alone
the so-called {'}t Hooft term~\cite{tHooft86} 
\begin{eqnarray}
{\cal L}&=&{\cal L}_{\mathrm{kinetic}}+{\cal L}_{\mathrm{four-fermion}}+
{\cal L}_{\mathrm{{'}t Hooft}}
\nonumber \\
{\cal L}_{\mathrm{{'}t Hooft}} &\sim& \mathrm{det}_f \left[
\bar{q} \left( 1 + \gamma_5 \right) q \right]
+ \mathrm{det}_f \left[\bar{q} \left( 1 - \gamma_5 \right) q
\right]
\label{eq:thooft}
\end{eqnarray}
is added where $q$ is a quark spinor and $\mathrm{det}_f$ is a
determinant in flavor space.  This term, like QCD, respects chiral
symmetry but breaks the $U(1)_A$ symmetry. The use of this Lagrangian
is the standard approach which has been used to describe dense matter.

Unfortunately, when employed to study quark superconductivity, 
this standard approach does not employ a manifestly
consistent truncation scheme; the quark--anti-quark interaction is
treated at the six-fermion level, but the quark-quark interaction is
only treated at the four-fermion level. There is no reason to 
rule out a term of the form (omitting color for clarity)
\begin{eqnarray}
{\cal L}_{\mathrm{CS6}} \sim K_{\mathrm{DIQ}} 
~\epsilon_{ijm} \epsilon_{k \ell n}~
({\bar q}_i \gamma_5 q_j^C) ({\bar q}_k^C \gamma_5 q_{\ell}) 
({\bar q}_m q_n)
\label{eq:newterm}
\end{eqnarray} 
which has the same symmetries as ${\cal L}_{\mathrm{{'}t Hooft}}$. This
term may have a significant impact on the nature of dense
matter~\cite{Rapp98,Schafer02,Buballa04}. While the effect of the
dynamically generated quark masses on the superconducting gaps has
been studied~\cite{Steiner02}, Eq. \ref{eq:newterm} implies a
modification to the quark masses due to the presence of the gap.

In this article, we study the effect of ${\cal L}_{\mathrm{CS6}}$ on
the quark masses and on the phase structure of dense matter.  We show
that sufficiently positive values of $K_{\mathrm{DIQ}}$ increase the
quark masses and thus favor less-gapped phases, while sufficiently
negative values of $K_{\mathrm{DIQ}}$ split the magnitude of the
up-down and light-strange gaps.  The phase structure of dense matter
thus depends critically on the sign and magnitude of this unknown
parameter.

\section{The Model Lagrangian}

The Lagrangian is~\cite{Buballa02,Gastineau02,Steiner02}
\begin{eqnarray}
{\cal L} &=& \bar{q}_{i \alpha} \left( i \feyn{\partial} - m_{0,i j} -
\mu^{i \alpha}_{j \beta} \gamma_0 \right) q_{j \beta} +
G \sum_{a=0}^8 \left( \bar{q} \lambda^a_f q \right)^2 \nonumber \\
&& + G_{\mathrm{DIQ}} \epsilon^{i j m} 
\epsilon^{k \ell m} 
\epsilon^{\alpha \beta \varepsilon} 
\epsilon^{\gamma \delta \varepsilon} 
\nonumber \\
&& \times
\left(\bar{q}_{i \alpha} i \gamma^5 
q^C_{j \beta}\right) \left(\bar{q}_{k \gamma}^C i \gamma^5
q_{\ell \delta}\right) \nonumber \\
&& - K \left[
\mathrm{det}_f \bar{q} \left( 1 - i \gamma^5 \right) q +
\mathrm{det}_f \bar{q} \left( 1 + i \gamma^5 \right) q\right] \nonumber \\
&& + K_{\mathrm{DIQ}}~\epsilon^{ijm} \epsilon^{k \ell n}
\epsilon^{\alpha \beta \varepsilon} \epsilon^{\gamma \delta \eta}
\nonumber \\
&& \times
({\bar q}_{i\alpha} i \gamma_5 q_{j\beta}^C)({\bar q}_{k\gamma}^C i \gamma_5 q_{\ell\delta})
({\bar q}_{m \varepsilon} q_{n \eta})
\label{colorlagr}
\end{eqnarray}
where flavor is represented by Latin indices, 
color is represented by Greek indices, 
and the charge conjugate Dirac spinors are defined by 
($C \gamma^{\mu} C = \gamma^{\mu T}$ and $C^T =- C$)
\begin{equation}
q^C \equiv C \bar{q}^{T} \quad \mathrm{and} \quad
\bar{q}^C = q^T C \, .
\end{equation}
The four-fermion coupling in the quark--anti-quark channel is denoted
$G$, the four-fermion coupling in the quark-quark channel is denoted
$G_{\mathrm{DIQ}}$, and $m_{0,ij}$ is the constant current quark mass
matrix which is diagonal in flavor.

We utilize the ans\"{a}tze
\begin{eqnarray}
{\bar q}_1 q_2 {\bar q}_3 q_4
&\rightarrow &
\left<{\bar q}_1 q_2\right> {\bar q}_3 q_4
+{\bar q}_1 q_2 \left<{\bar q}_3 q_4 \right>
-\left<{\bar q}_1 q_2\right>\left< {\bar q}_3 q_4 \right>
\nonumber \\
{\bar q}_1 q_2 {\bar q}_3 q_4 {\bar q}_5 q_6
&\rightarrow &
{\bar q}_1 q_2 \left<{\bar q}_3 q_4\right> \left<{\bar q}_5 q_6\right>
+\left<{\bar q}_1 q_2\right> \left<{\bar q}_3 q_4\right> {\bar q}_5 q_6
\nonumber \\ &&
+\left<{\bar q}_1 q_2\right> {\bar q}_3 q_4 \left<{\bar q}_5 q_6\right>
\nonumber \\ &&
-2 \left<{\bar q}_1 q_2\right> \left<{\bar q}_3 q_4\right>
\left<{\bar q}_5 q_6\right>
\label{eq:ansatz}
\end{eqnarray}
to obtain the mean-field approximation~\cite{Bernard88}.  This
procedure retains only the lowest order terms in the $1/N_c$
expansion~\cite{Klevansky92}. Color neutrality is ensured using the
procedure from Ref.~\cite{Steiner02}. In the mean-field approximation,
the Lagrangian is only quadratic in the fermion fields, and the
thermodynamical potential can be obtained from the inverse propagator
in the standard way \cite{Kapusta85}.  The momentum integrals in the
gap equations are divergent, and are regulated by a three-momentum
cutoff denoted by $\Lambda$. The inverse propagator is numerically
diagonalized for each abscissa of the momentum integration to obtain
the thermodynamical potential.
 
For simplicity, we sometimes use the notation $\Delta_k \sim
\epsilon^{ijk} \left<q^C_i \gamma^5 q_j \right>$, so that gaps are
denoted with the flavor of quark that is {\it not} involved in the
pairing e.g. $\Delta_{ud}$ is denoted by $\Delta_s$. Other than
$\left<{\bar q}_i q_i \right>$ and $\Delta_i$, we assume that all
other condensates vanish. This includes the pseudoscalar condensates
which which are likely present in dense matter and naturally accompany
the Goldstone bosons~\cite{Kaplan02}. This (non-trivial) complication
will be left to later work.

The effects of ${\cal L}_{\mathrm{CS6}}$ on the thermodynamical
potential can be summarized in three modifications from the standard
approach where $K_{\mathrm{DIQ}}=0$. These changes are that the values
of the gap in the inverse propagator are modified
\begin{eqnarray}
\Delta_i \rightarrow \Delta_i \left(1 + \frac{K_{\mathrm{DIQ}}}
{N_c G_{\mathrm{DIQ}}}
\left< {\bar q}_i q_i \right> \right) \, ,
\label{eq:modfirst}
\end{eqnarray}
a new effective mass term (which includes contributions
which are not diagonal in flavor) appears
\begin{eqnarray}
\frac{K_{\mathrm{DIQ}}}{4 G_{\mathrm{DIQ}}^2} 
{\bar q}_i \Delta_i \Delta_j q_j \, ,
\label{eq:effmass}
\end{eqnarray}
which modifies the dynamical mass
\begin{eqnarray}
m^{*}_i=m_{i,0}-4 G \left<\bar{q} q\right>_i+K |\epsilon_{i j k}|
\left<\bar{q} q\right>_j \left<\bar{q} q\right>_k+
\frac{K_{\mathrm{DIQ}}}{4 G_{\mathrm{DIQ}}^2} \Delta_i^2
\label{eq:effmass2}
\end{eqnarray}
and that there is a new contribution to the vacuum energy
\begin{eqnarray}
\Omega_{K_{\mathrm{DIQ}}} = \frac{K_{\mathrm{DIQ}}}{2
G_{\mathrm{DIQ}}^2} \sum_i \Delta_i^2 
\left< {\bar q}_i q_i \right> \, .
\label{eq:modlast}
\end{eqnarray}
Note that Eq. \ref{eq:effmass} means that the quark masses are
density-dependent if $K_{\mathrm{DIQ}}\neq 0$ even when the chiral
condensates $\left<{\bar q} q \right>$ vanish. As has been
suggested~\cite{Rajagopal00}, this term generates a dynamical quark
mass entirely distinct from the typical mechanism of spontaneous
chiral symmetry breaking. The quarks obtain a dynamical mass even when
the quark condensate is taken to be zero.

One may use Fierz transformations to calculate the coefficient
$K_{\mathrm{DIQ}}$ from the quark--anti-quark form of the 
{'}t Hooft interaction. One can view this in
the following way: For each prospective new term, e.g. the term ${\bar
u} \gamma^5 d_c {\bar d}_c \gamma^5 u {\bar s} s$, there are six Fierz
transformations (for six fermions this is a 35 $\times$ 35 matrix
instead of the usual 5 $\times$ 5 matrix for four fermions)
that give this term when applied to the six sets (in flavor space) of
terms in the {'}t Hooft interaction. One such transformation is
\begin{eqnarray}
&& {\bar q}_i q_j {\bar q}_k q_{\ell} {\bar q}_m q_n 
+{\bar q}_i \gamma^5 q_j {\bar q}_k \gamma^5 q_{\ell} {\bar q}_m q_n 
\nonumber \\
&& +{\bar q}_i q_j {\bar q}_k \gamma^5 q_{\ell} {\bar q}_m \gamma^5 q_n 
+{\bar q}_i \gamma^5 q_j {\bar q}_k q_{\ell} {\bar q}_m \gamma^5 q_n 
\nonumber \\
&=& 
\frac{1}{2} \left(
{\bar q}_i \gamma^5 q_{k}^C {\bar q}_{j}^C \gamma^5 q_{\ell} 
{\bar q}_m q_n
+{\bar q}_i q_{k}^C {\bar q}_{j}^C \gamma^5 q_{\ell} 
{\bar q}_m \gamma^5 q_n \right.
\nonumber \\ &&
\left. +{\bar q}_i q_{k}^C \gamma^5 {\bar q}_{j}^C q_{\ell} 
{\bar q}_m \gamma^5 q_n
+{\bar q}_i q_{k}^C {\bar q}_{j}^C q_{\ell} {\bar q}_m q_n
\right)
\nonumber \\
&&- \frac{1}{4} \left(
{\bar q}_i \gamma^5 \sigma^{\mu \nu} q_{k}^C 
{\bar q}_{j}^C \gamma^5 \sigma_{\mu \nu} q_{\ell} {\bar q}_m q_n
\right. \nonumber \\ && \left.
+{\bar q}_i \gamma^5 \sigma^{\mu \nu} q_{k}^C 
{\bar q}_{j}^C \sigma_{\mu \nu} q_{\ell} {\bar q}_m \gamma^5 q_n
\right)
\label{eq:fex}
\end{eqnarray}
When these transformations are combined to give the coefficient
$K_{\mathrm{DIQ}}$, the result is zero (see the Appendix). Although
terms with different Dirac structure do survive, we do not include
these terms since we do not expect the corresponding condensates,
e.g. $\left<\bar{q} \sigma_{\mu \nu} q\right>$ to be non-zero. This
procedure is not the only possible approach for deriving the
mean-field Lagrangian (one could also enumerate all possible Wick
contractions). Because alternate approaches and/or using terms of
higher order in $1/N_c$ may modify this result, we cannot conclude
necessarily that $K_{\mathrm{DIQ}}$ must be zero. Also, terms like
Eq. \ref{eq:thooft} with a Dirac structure $(\bar{q} q) (\bar{q}
\sigma_{\mu \nu} q) (\bar{q} \sigma^{\mu \nu} q)$ \cite{Shifman80} and
their corresponding terms in the quark-quark channel may play a
role. We leave these considerations to future work.

Our Lagrangian is free to contain any terms which follow the
symmetries of the underlying theory. We expect that the coefficient of
this term will be ``natural''. When the coefficients are expressed in
terms of the underlying length scales (in our case, the momentum
cutoff $\Lambda$), the coefficients should all be of similar
magnitude. We allow the coefficient $K_{\mathrm{DIQ}}$ to vary,
between the values $-K$ and $K$, which we view to a modest variation
as suggested by the constraints of naturalness. A larger variation in
$K_{\mathrm{DIQ}}$ is not necessarily excluded. We use the values of
$\Lambda$ and the current quark masses from Ref.~\cite{Rehberg96},
where they are fixed by matching the pion, kaon, and $\eta^{\prime}$
masses in vacuum as well as the pion decay constant. We choose to fix
$G_{\mathrm{DIQ}} \Lambda^2=1.61$ to be large enough so that the
maximum value of the gaps (when including the quark dynamical mass) as
a function of density when $K_{\mathrm{DIQ}}=0$ is about 80 MeV, close
to the value of about 100 MeV predicted by calculations in
perturbative QCD. If the dynamically-generated quark mass is assumed
to be zero and the mass-gap equations are ignored, then the maximum
value of the gap predicted by this model is about 120 MeV. We leave
$G_{\mathrm{DIQ}}$ fixed when varying $K_{\mathrm{DIQ}}$. One could
also allow the coefficient, $G_{\mathrm{DIQ}}$ as a function to vary
as a function of $K_{\mathrm{DIQ}}$ by instead ensuring that the
maximum value of one of the three gaps at high densities is constant.
We have checked that this alternative does not change our conclusions
significantly.

\section{Results}

Obtaining analytical results is difficult, due to the flavor mixing
mass terms from Eq. \ref{eq:effmass} which make it difficult to
directly reduce the inverse propagator (a 36 $\times$ 36 matrix) into
a block-diagonal form. It is possible, with High-Density Effective
Theory~\cite{Hong00}, to simplify the Dirac structure, but this
would likely result in a 9 $\times$ 9 inverse propagator which is also
difficult. It is also possible to restore the usual form of the
propagator encountered in studies where $K_{\mathrm{DIQ}}=0$ by
ignoring the terms in Eq. \ref{eq:effmass} where $i\neq j$. In this
case, the inverse propagator is worked out in detail in
Ref.~\cite{Ruster05}.

It is possible to see qualitatively, what the effect of adding a term
with $K_{\mathrm{DIQ}} \neq 0$ might be from Eq. \ref{eq:modfirst}
above. When $K_{\mathrm{DIQ}}>0$, we expect the gaps decrease as the
quark condensate increases, and thus the gap should decrease with
increasing mass. However, from Eq. \ref{eq:effmass2} we expect the
opposite and we find that it is this effect that dominates the
description of the strange quark mass and $\Delta_{ud}$.  Further
complicating the analysis, Eq. \ref{eq:effmass} indicates that an
increase in $\Delta_{us}$ and $\Delta_{ds}$ will tend to split the
mass of the up and down quark, thus possibly weakening $\Delta_{ud}$.

We study charge- and color-neutral, beta-equilibrated, bulk matter at
fixed baryon density and a fixed temperature of $10$ MeV. We operate
at a small but finite temperature in order to alleviate the numerical
difficulties of discontinuities in the momentum integral present in
the thermodynamical potential. The zero-temperature results will not
deviate significantly from our results. We include non-interacting
electrons, but we do not include neutrinos. The addition of neutrinos
would further split the approximate flavor symmetry between the up and
down quarks. Our results will faithfully describe matter in the center
of a neutron star containing quarks a minute or later after
formation~\cite{Prakash95,Prakash97}.

We note that the effect of $K_{\mathrm{DIQ}}$ is small when the quark
condensates $\left< \bar{q} q\right>$ are taken to be zero. When this
is assumed to be true, then the modifications from
Eqs. \ref{eq:modfirst} and \ref{eq:effmass} have no effect, and the
gaps are nearly independent of $K_{\mathrm{DIQ}}$. We remove this
assumption and solve the mass gap equations for the quark condensates
in the following.

\begin{figure}[htb]
\begin{center}
\includegraphics[scale=0.42]{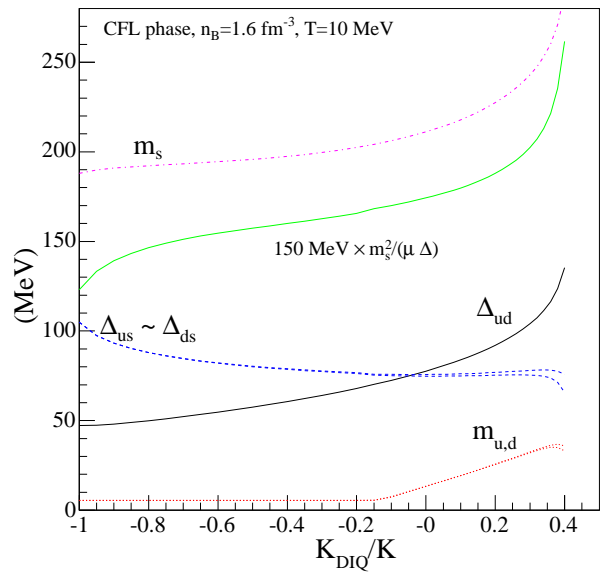}
\end{center}
\caption{Quark masses (dotted lines for u and d, and dashed-dotted
line for s) and gaps (solid line for $\Delta_{ud}$ and dashed line for
the other two gaps) as a function of $K_{\mathrm{DIQ}}$ in the CFL
phase at $n_B=1.6$ fm$^{-3}$.}
\label{fig:16sixdm}
\end{figure}

Figure \ref{fig:16sixdm} presents the masses and gaps in the CFL phase
at fixed density and temperature as a function of $K_{\mathrm{DIQ}}$.
Both the quark masses and $\Delta_{ud}$ increase as $K_{\mathrm{DIQ}}$
increases. The effect from Eq. \ref{eq:effmass} causes the
$\Delta_{us}$ and $\Delta_{ds}$ gap to decrease when $\Delta_{ud}$
increases. At sufficiently large values of the coupling, the strong
increase of the strange quark mass destabilizes the CFL phase. For
$K_{\mathrm{DIQ}}>0.4$, the gap equations have no solution. If the
coupling was verified by some other means to be larger than this
critical value, then the CFL phase could not be present at this
density. As $K_{\mathrm{DIQ}}/K \rightarrow -1$, the effects on the
masses and gaps tend to be less extreme. The most significant effect
is the increasing split between the values of the light-strange gaps,
$\Delta_{us}$ and $\Delta_{ds}$, and the light-quark gap
$\Delta_{ud}$. One might expect the dependence of the gaps on
$K_{\mathrm{DIQ}}$ would change the phase structure of matter by
shifting the energy density. However, we find that this is not the
case and that the energy density is relatively constant as a function
of $K_{\mathrm{DIQ}}$. Note also that the strange quark mass can
change by as much as 50\% for different values of $K_{\mathrm{DIQ}}$.

Also plotted in Fig.~\ref{fig:16sixdm} is the parameter $m_s^2/(\mu
\Delta)$ where $\mu$ and $\Delta$ in this context are computed by
averaging the quark chemical potentials and gaps over all flavors and
colors. This parameter has been demonstrated to be the relevant
dimensionless quantity which dictates the phase content of quark
matter at high density~\cite{Alford01}. Values of $m_s^2/(\mu \Delta)$
larger than 2 suggest a transition to a gapless CFL phase~\cite{Alford05},
while values larger than 4 suggest a transition to the 2SC phase. 
The presence of the gapless 2SC phase~\cite{Shovkovy03} is also 
probable when $K_{\mathrm{DIQ}}$ becomes positive, since the $ud$ gap
is becoming stronger and the light-strange gaps are weakening. The
value of $m_s^2/(\mu \Delta)$ is also important in dictating the
number and type of Goldstone bosons present in the CFL phase.  This
parameter is quite flat for small variations in $K_{\mathrm{DIQ}}$,
but increases or decreases strongly when the absolute magnitude of
$K_{\mathrm{DIQ}}$ is sufficiently large. When $K_{\mathrm{DIQ}}/K\sim
0.4$, the value of $m_s^2/(\mu \Delta)$ is nearly 2, indicating the
disappearance of the CFL phase.

\begin{figure}[htb]
\begin{center}
\includegraphics[scale=0.42]{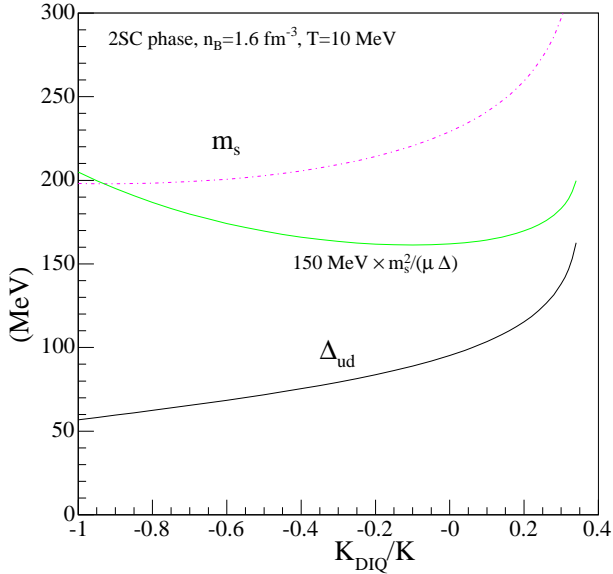}
\end{center}
\caption{Strange quark mass (dashed-dotted line) and the gap (solid
line) as a function of $K_{\mathrm{DIQ}}$ in the 2SC phase at
$n_B=1.6$ fm$^{-3}$. }
\label{fig:16six2dm}
\end{figure}

The results for the 2SC phase, where $\Delta_{us}=\Delta_{ds}=0$, at
the same density and temperature are shown in Figure
\ref{fig:16six2dm}. The strange quark mass and gap increase strongly
for increasing $K_{\mathrm{DIQ}}$ as in the CFL phase, leading to a
critical value above which the gap equations have no solution. At this
density, the 2SC phase is also not present in matter for
$K_{\mathrm{DIQ}}>0.4$. The similarity of this critical value of the
coupling to the CFL phase in Fig.~\ref{fig:16sixdm} is a result of the
fact that the light-strange gaps and light-quark masses are comparatively
small and thus do not strongly perturb $m_s$ and $\Delta_{ud}$. The
parameter $m_s^2/(\mu \Delta)$ is slightly modified from the CFL
case and indicates that the 2SC phase is also likely to be unstable
for large negative values of $K_{\mathrm{DIQ}}/K$.

\begin{figure}[htb]
\begin{center}
\includegraphics[scale=0.42]{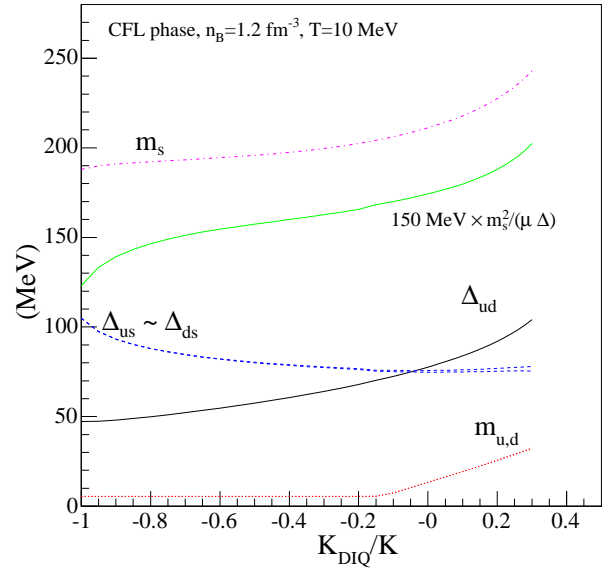}
\end{center}
\caption{The quark masses and gaps in the CFL phase at 
$n_B=1.2$ fm$^{-3}$. See Fig.~\ref{fig:16sixdm}.}
\label{12sixdm}
\end{figure}

For comparison, Fig.~\ref{12sixdm} presents the quark masses and gaps
in the CFL phase at a lower density, 1.2 fm$^{-3}$. The results are 
not much different from Fig.~\ref{fig:16sixdm}. The quark gaps have
decreased slightly and the quark masses are slightly larger. The major
distinction is that the critical value of $K_{\mathrm{DIQ}}$ above
which the CFL phase does not appear has been lowered from 0.4 to 
less than 0.3.

\begin{figure}[htb]
\begin{center}
\includegraphics[scale=0.42]{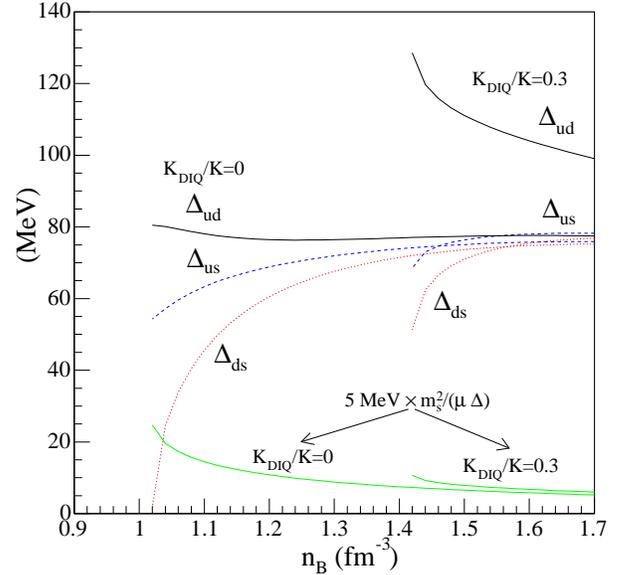}
\end{center}
\caption{The gaps as a function of baryon density in the CFL
phase for two values of the coupling $K_{\mathrm{DIQ}}$. }
\label{sixdm_tr}
\end{figure}

The implication of the shift in the critical value of
$K_{\mathrm{DIQ}}$ is that the critical density for the onset of the
CFL phase is drastically affected by a positive value of
$K_{\mathrm{DIQ}}$. This is demonstrated in Fig.~\ref{sixdm_tr}, where
the smallest quark gap vanishes at about 1.0 fm$^{-3}$ when
$K_{\mathrm{DIQ}}=0$, and at about 1.42 fm$^{-3}$ when
$K_{\mathrm{DIQ}}/K=0.3$. On the other hand, because the gaps are not
as strongly modified when $K_{\mathrm{DIQ}}$ is negative, the critical
density when $K_{\mathrm{DIQ}}/K=-0.5$ is almost unchanged, moving
down only to 0.96 fm$^{-3}$. In this figure, the $\Delta_{ds}$ gap
does not vanish completely to zero because the solution of the gap
equations becomes difficult when the gaps are small. Note again that
an increase the parameter $m_s^2/(\mu \Delta)$ indicates, to some
extent, the disappearance of the CFL phase as the density decreases.

\section{Discussion - Neutron Stars with $K_{DIQ} \neq 0$} 

These results may have several implications for neutron star structure
and evolution. 

\begin{figure}[htb]
\begin{center}
\includegraphics[scale=0.42]{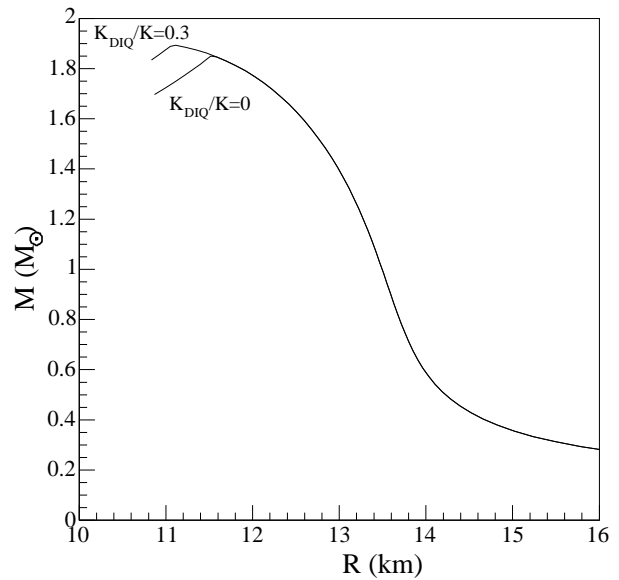}
\end{center}
\caption{The mass versus radius for neutron stars with two values of
the coupling $K_{\mathrm{DIQ}}$. }
\label{fig:mvsr}
\end{figure}

{\it Neutron-star masses and radii}: We have solved the
Tolman-Oppenheimer-Volkov equations for neutron star structure under
the assumption that there is no mixed phase (i.e. the surface
tension is very large so that mixed phases are suppressed). The
results for $K_{\mathrm{DIQ}}=0$ and $K_{\mathrm{DIQ}}/K=0.3$ are
given in Fig.~\ref{fig:mvsr}. We find that neutron star masses and
radii are not very sensitive to $K_{\mathrm{DIQ}}$ for the model that
we have chosen. As mentioned above, a positive value of
$K_{\mathrm{DIQ}}$ tends to increase the critical density for the
appearance of the CFL phase, thus stiffening the equation of state and
slightly increasing the maximum mass from 1.83 to 1.9 $M_{\odot}$.
The small magnitude of this effect is partially due to the fact that,
for $K_{DIQ}=0$, the critical density for the appearance of quark
matter is 1.0 fm$^{-3}$, while the central density of the maximum mass
neutron star is only 1.45 fm$^{-3}$.  There is not much quark matter
to begin with, so therefore the effect of $K_{\mathrm{DIQ}}$ is
limited. In regards to the phase content of matter in the neutron
star, the larger value of $K_{\mathrm{DIQ}}$ nearly pushes the CFL
phase out of the neutron star entirely, and the center of the neutron
star is dominated by the 2SC phase instead.  These effects will be
larger if the diquark coupling is increased and may be modified by the
presence of a mixed phase. Recent calculations of the surface tension
suggest that it is small, and thus a mixed phase may be
present~\cite{Reddy05}. It would be interesting to examine the effect
of $K_{\mathrm{DIQ}}$ on this surface tension.

{\it Neutron-star cooling}: Neutron stars are sensitive to the
difference between the CFL and 2SC phases since the former is likely
to contain no~\cite{Rajagopal01} (or very few) electrons. Neutron star
cooling is affected by the presence or absence of electrons because
the specific heat of matter is dominated by the electrons when they
are present. In addition, the specific heat contribution from the
light quarks is proportional to $\exp(-\Delta_{\ell s}/T)$, where
$\ell=u~\mathrm{or}~d$ which is exponentially small in the CFL phase
and of order unity in the 2SC phase. Also, the splitting of the gaps at
$K_{\mathrm{DIQ}}/K=-1$, will enforce two critical temperatures for
quark matter in the CFL phase. The so-called ``gapless'' phases
are dependent upon the strange quark mass which is strongly affected
by $K_{\mathrm{DIQ}}$, especially when it is greater than zero. These
gapless phases, because of the nature of the quark dispersion
relations, have unusual transport properties that also have
implications for neutron star cooling~\cite{Shovkovy03,Shovkovy04,Alford05}.

{\it Proto-neutron star evolution}: Ref.~\cite{Carter00} pointed out
that the proto-neutron star neutrino signal may be increased noticeably
by the enhanced cooling that is present when the core temperature
falls below the critical temperature. The presence of
$K_{\mathrm{DIQ}}$ implies that this enhanced cooling may occur in (at
least) two separate stages, as the critical temperatures corresponding
to the $\Delta_{ud}$ and the light-strange gaps are surpassed.

To the extent that the presence of a large ($\sim 100$ MeV) gap
affects observations of neutron stars, the presence of a
color-superconducting six-fermion interaction also has an impact on
neutron star observations. It would be interesting to compare these
results with those from gapless CFL phases.  Because of the
uncertainty in the value of the $K_{\mathrm{DIQ}}$ coupling
(in addition to the other uncertainties already present), it will
be difficult to settle the questions of the nature of dense matter and
the question of the properties of neutron stars containing deconfined
quark matter until theoretical progress in QCD or astrophysical
observations can more completely settle the issue.

The author would like to thank T. Bhattacharaya, T. B\"{u}rvenich,
M. Prakash, S. Reddy, G. Rupak, and T. Sch\"{a}fer for discussions.
This work has been supported by the DOE under grant number
DOE/W-7405-ENG-36.

\section*{Appendix A - Fierz Transformations}

Traditionally, the Fierz transformation is defined as the matrix, $C$,
which obeys the relation
\begin{eqnarray}
{\cal F}\left[
\bar{\psi}_{i} \Gamma_{a,ij} \psi_j 
\bar{\psi}_k \Gamma^{a}_{k \ell} \psi_\ell 
\right] =
\sum_b C_{a,b}
\bar{\psi}_{i} \Gamma_{b,i \ell} \psi_\ell
\bar{\psi}_k \Gamma^{b}_{k j} \psi_j \,
\end{eqnarray}
where $\Gamma_i \in \left[\bfone,\gamma^5,\gamma_{\mu}, \gamma^5
\gamma_{\mu},\sigma_{\mu \nu} \right]$ and $\Gamma^i \in
\left[\bfone,\gamma^5,\gamma^{\mu}, \gamma^5 \gamma^{\mu},\sigma^{\mu
\nu} \right]$ for $i=a,b$.  Because the Fierz transformation is
nothing other than a set of equalities, the two four-fermion
interactions on both sides are necessarily equivalent. However, in the
mean-field approximation, these two forms lead to different
thermodynamic potentials. For simplicity, the transformation above
will be denoted ${\cal F}(ijk \ell \rightarrow i \ell k j)$.

One may also perform a four-fermion Fierz transformation in the
quark-quark channel, namely, ${\cal F}(ijk \ell \rightarrow i k
j \ell)$ (see the review in Ref.~\cite{Nieves04}). 
This transformation operates over a different ``basis'' of
combinations of Dirac matrices
\begin{eqnarray}
{\cal F}\left[
\bar{\psi}_{i} \Gamma_{a,ij} \psi_j 
\bar{\psi}_k \Gamma^{a}_{k \ell} \psi_\ell 
\right] = \qquad \qquad\qquad\qquad \nonumber \\
\qquad\qquad\qquad\qquad \sum_b C^{\prime}_{a,b}
\bar{\psi}_{i} \Gamma^{\prime}_{b,i k} \psi_k^C
\bar{\psi}_j^C \Gamma^{\prime,b}_{j \ell} \psi_{\ell} \, .
\end{eqnarray}
where $\Gamma^{\prime}_i \in \left[\gamma_5 ,\bfone ,\gamma^5
\gamma_{\mu}, \gamma_{\mu}, \sigma_{\mu \nu} \right]$ and
$\Gamma^{i,\prime} \in \left[\gamma^5,\bfone, \gamma^5 \gamma^{\mu},
\gamma^{\mu}, \sigma^{\mu \nu} \right]$ (remember that $\bar{\psi}^C
\gamma^5 \psi$ is a Lorentz scalar). We can simplify the notation for
the basis by using the notation of a direct product: $\gamma^5 \otimes
\gamma^5, \bfone \otimes \bfone, \gamma^5 \gamma_{\mu} \otimes
\gamma^5 \gamma_{\mu}, \gamma_{\mu} \otimes \gamma_{\mu}, \sigma_{\mu
\nu} \otimes \sigma^{\mu \nu}$, or more simply, $\mathrm{S S, P P, V
V, A A, T T}$.

In a similar notation the 35-element basis for the six-fermion Fierz
transformations is
\begin{eqnarray}
&\mathrm{SSS,SPP,PSP,PPS,SVV,VSV,VVS,}& \nonumber \\
&\mathrm{SAA,ASA,AAS,STT,TST,TTS,}& \nonumber \\
&\mathrm{PVA,PAV,VPA,VAP,APV,AVP,}& \nonumber \\
&\mathrm{TVV,VTV,VVT,TAA,ATA,AAT,TTT,}& \nonumber \\
&\mathrm{VAQ,VQA,AVQ,AQV,QVA,QAV,}& \nonumber \\
&\mathrm{PTQ,TPQ,QTP}&
\end{eqnarray}
where $Q$ denotes a ``pseudo-tensor'' combination, $\gamma^{5}
\sigma^{\mu \nu}$. (The $\mathrm{Q}$ terms are used as an alternative
to the formulation in terms of objects of the form $\varepsilon_{\kappa
\lambda \mu \nu} \sigma^{\mu \nu}$~\cite{Maruhn01}.) All other
combinations can be written as a linear combination of these 35 basis
elements. In order to distinguish 4- and 6-fermion transformations, we
will use a subscript, i.e. ${\cal F}_4 (ijk\ell mn \rightarrow i \ell
k j m n)$ is really a four-fermion transformation since the fields
with indices $m$ and $n$ are not participating in the transformation.

The Dirac scalar terms in the {'}t Hooft interaction are
\begin{eqnarray}
& \bar{u} u \bar{d} d \bar{s} s
+\bar{u} s \bar{d} u \bar{s} d
+\bar{u} d \bar{d} s \bar{s} u 
& \nonumber \\ &
-\bar{u} d \bar{d} u \bar{s} s
-\bar{u} s \bar{d} d \bar{s} u
-\bar{u} u \bar{d} s \bar{s} d
& \, .
\end{eqnarray}
plus the corresponding terms created by adding an even number of
$\gamma^5$ matrices. 

As a demonstration, we examine the coefficient of the term $\bar{u}
\gamma^5 d^C \bar{u}^C \gamma^5 d \bar{s} s$. 
The various contributions to this coefficient are
\begin{eqnarray}
& +{\cal F}_4(ijklmn \rightarrow ikjlmn)
& \left[ 
\bar{u} u \bar{d} d \bar{s} s
+ \bar{u} u \bar{d} \gamma^5 d \bar{s} \gamma^5 s
\right. \nonumber \\ && \left.
+ \bar{u} \gamma^5 u \bar{d} d \bar{s} \gamma^5 s
+ \bar{u} \gamma^5 u \bar{d} \gamma^5 d \bar{s} s
\right] \nonumber \\
& + {\cal F}_6(ijklmn \rightarrow iklnmj)
& \left[
\bar{u} s \bar{d} u \bar{s} d
+ \bar{u} s \bar{d} \gamma^5 u \bar{s} \gamma^5 d
\right. \nonumber \\ && \left.
+ \bar{u} \gamma^5 s \bar{d} u \bar{s} \gamma^5 d
+ \bar{u} \gamma^5 s \bar{d} \gamma^5 u \bar{s} d
\right] \nonumber \\
& + {\cal F}_6(ijklmn \rightarrow iknjml)
&\left[
\bar{u} d \bar{d} s \bar{s} u
+ \bar{u} d \bar{d} \gamma^5 s \bar{s} \gamma^5 u
\right. \nonumber \\ && \left.
+ \bar{u} \gamma^5 d \bar{d} s \bar{s} \gamma^5 u
+ \bar{u} \gamma^5 d \bar{d} \gamma^5 s \bar{s} u
\right] \nonumber \\
& - {\cal F}_4(ijklmn \rightarrow ikljmn)
&\left[
\bar{u} d \bar{d} u \bar{s} s
+ \bar{u} d \bar{d} \gamma^5 u \bar{s} \gamma^5 s
\right. \nonumber \\ && \left.
+ \bar{u} \gamma^5 d \bar{d} u \bar{s} \gamma^5 s
+ \bar{u} \gamma^5 d \bar{d} \gamma^5 u \bar{s} s
\right] \nonumber \\
& - {\cal F}_6(ijklmn \rightarrow iknlmj)
&\left[
\bar{u} s \bar{d} d \bar{s} u
+ \bar{u} s \bar{d} \gamma^5 d \bar{s} \gamma^5 u
\right. \nonumber \\ && \left.
+ \bar{u} \gamma^5 s \bar{d} d \bar{s} \gamma^5 u
+ \bar{u} \gamma^5 s \bar{d} \gamma^5 d \bar{s} u
\right] \nonumber \\
& - {\cal F}_6(ijklmn \rightarrow ikjnml)
&\left[
\bar{u} u \bar{d} s \bar{s} d
+ \bar{u} u \bar{d} \gamma^5 s \bar{s} \gamma^5 d
\right. \nonumber \\ && \left.
+ \bar{u} \gamma^5 u \bar{d} s \bar{s} \gamma^5 d
+ \bar{u} \gamma^5 u \bar{d} \gamma^5 s \bar{s} d
\right] \nonumber
\end{eqnarray}
Note that result of the first of these six transformations is given as
Eq.~\ref{eq:fex} in the text. The coefficients of the desired term,
$\bar{u}\gamma^5 d^C \bar{u}^C \gamma^5 d \bar{s} s$,
from each of the six transformations (together with an 
appropriate factor of -1 for odd fermionic permutations) are
\begin{equation}
\textstyle{\frac{1}{2}},-\textstyle{\frac{1}{4}},-\textstyle{\frac{1}{4}},
-\textstyle{\frac{1}{2}},\textstyle{\frac{1}{4}},\textstyle{\frac{1}{4}}
\end{equation}
and the sum is zero.

Because the coefficient is zero, we need not consider the Fierz
transformations in the SU(3) (color or flavor) spaces. However, since
the result for six-fermion transformations in SU(3) is not present in
the literature, we give the 15-element basis for computing the
transformations (this enlarged basis from that presented in
Ref.~\cite{Dmitrasinovic01} is necessary to perform the
transformations in the quark--quark channel)
\begin{eqnarray}
& \bfone \otimes \bfone \otimes \bfone, \qquad
\lambda^a \otimes \lambda_a \otimes \bfone, & \nonumber \\
& \bfone \otimes \lambda^a \otimes \lambda_a, \qquad 
\lambda_a \otimes \bfone \otimes \lambda^a, & \nonumber \\
& d_{abc} \lambda^a \otimes \lambda^b \otimes \lambda^c, \qquad 
i f_{abc} \lambda^a \otimes \lambda^b \otimes \lambda^c, & \nonumber \\
& \lambda(A)^a \otimes \lambda(A)_a \otimes \bfone, \quad 
\bfone \otimes \lambda(A)^a \otimes \lambda(A)_a, & \nonumber \\
& \lambda(A)_a \otimes \bfone \otimes \lambda(A)^a,
& \nonumber \\
& \lambda(A)_a \lambda(S)_b \otimes \lambda(A)^a \otimes \lambda(S)^b,
& \nonumber \\
& \lambda(A)_a \lambda(S)_b \otimes \lambda(S)^b \otimes \lambda(A)^a,
& \nonumber \\
& \lambda(S)^b \otimes \lambda(A)_a \lambda(S)_b \otimes \lambda(A)^a,
& \nonumber \\
& \lambda(A)^a \otimes \lambda(A)_a \lambda(S)_b \otimes \lambda(S)^b,
& \nonumber \\
& \lambda(A)^a \otimes \lambda(S)^b \otimes \lambda(A)_a \lambda(S)_b,
& \nonumber \\
& \lambda(S)^b \otimes \lambda(A)^a \otimes \lambda(A)_a \lambda(S)_b &
\end{eqnarray}
where the occurrence of $\lambda$ indicates an implicit sum over all 8
SU(3) matrices, $\lambda(A)$ restricts the sum to only the three
anti-symmetric $\lambda$ matrices, and the implicit sum for
$\lambda(S)$ is over the six symmetric $\lambda$ matrices. The full
results are available from the author.

\bibliographystyle{apsrev}
\bibliography{paper8}
\end{document}